\documentclass[
 amsmath,amssymb,
 aps,
]{revtex4-1}

\usepackage{graphicx}
\usepackage{dcolumn}
\usepackage{bm}
\usepackage[T1]{fontenc}       
\usepackage{amsfonts}
\usepackage{hyperref}
\usepackage{siunitx}

\begin{document}

\title[Oil Mass Exchanges in Shallow Waters]
  {Mass Exchange Dynamics of Surface and Subsurface Oil in Shallow-Water Transport}
  
\author{Saeed Moghimi}
\affiliation{Department of Civil Engineering, Portland State University, Portland, OR, U.S.A.}
\author{Jorge Ram{\'i}rez}
\affiliation{Departamento de Matem{\'a}ticas, Universidad de Colombia, Sede Medell{\'i}n, Medell{\'i}n Colombia}
\author{Juan M. Restrepo}
\email{restrepo@math.oregonstate.edu}
\affiliation{Department of Mathematics, Oregon State University, Corvallis OR, U.S.A.}
\altaffiliation{College of Earth, Ocean and Atmospheric  Sciences, Corvallis OR, U.S.A.}
\author{Shankar Venkataramani}
\affiliation{Mathematics Department, University of Arizona, Tucson, AZ 85721 U.S.A.}

\begin{abstract}
We  formulate  a model for the mass exchange between oil at and below the sea surface. This is a particularly important aspect of modeling oil spills. Surface and subsurface oil have  different  chemical and transport characteristics and lumping them together would compromise the accuracy of the resulting model. Without observational or computational constraints, 
 it is thus not possible to quantitatively predict oil spills based upon partial field observations of surface and/or sub-surface oil.
 The primary challenge in capturing the mass exchange  is that the principal mechanisms are on the microscale. This is a serious barrier to developing practical models for oil spills that are capable of addressing questions regarding  the  fate of oil at the large spatio-temporal scales, as demanded by environmental questions. We use upscaling to propose an environmental-scale
 model which incorporates the mass exchange between surface and subsurface oil due to oil droplet dynamics, buoyancy effects, and sea surface and subsurface mechanics.  
While the mass exchange mechanism   detailed here is generally applicable to oil transport models, it addresses the modeling needs of a particular to an oil spill model\cite{oil1}. This transport model is designed to capture oil spills at very large spatio-temporal scales. It accomplishes this goal by
specializing to shallow-water environments, in which  depth averaging is a perfectly good approximation for the flow, while at the same time
retaining mass conservation of oil over the whole oceanic domain.
\end{abstract}






\maketitle

\section{Introduction}
\label{intro}

A critical component of a mass-conserving oil fate model for oceanic oil spills is a model of the mechanics of oil exchange
between surface oil and sub-surface oil. Our primary focus in this work is on capturing the interplay between buoyancy, shearing and advection, as they affect the overall balance of oil in the surface and in the subsurface. In order to do so one has to include many of the details of the microscopic dynamics, for instance the changes in surface tension and in viscosity over time as a result of chemical reactions.  In particular, it is very important to track the droplet size distribution, since the size governs how micro-scale processes like merging, diffusion/dispersion, advection, shearing and buoyancy affect a dense population of droplets. 

Our focus is on oil transport and spills at environmental scales ({\it e.g.}, at the small end these would be kilometers and hours, at the larger scales it would be estuarine/basin scales and months/years). Our long-term goal is to develop a model for oil spill dynamics that could be implemented as a simulation platform. This is a daunting challenge for the following reasons:
\begin{enumerate}
\item The environmental scales are several orders of magnitude larger than the microscopic scales characterizing of  the droplet dynamics alluded to previously. A model that resolves the dynamics on the microscopic scale will end up with an enormous number of degrees of freedom.
\item  Oil is a composite, consisting of hundreds or thousands of different chemical species. This further compounds the ``curse of dimensionality'', if one seeks to accurately capture the multitude of chemical reactions within the oil. An alternative is to lump the individual species into a few chemical complexes. This approach now requires modeling the important chemical processes by empirical rules rather than from first principles. 
\item Finally, there is the matter of coupling the oil transport to models for the ocean, the atmosphere, and land.
\end{enumerate}   

In order to reach environmental scales we have embarked on the development of an oil transport model that was first introduced by Restrepo, {\it et al} \cite{oil1} (hereon, RRV). In developing this model, we have placed an emphasis respecting the underlying physics and in reducing its overall dimensionality, possibly at the expense of its quantitative accuracy. The rationale for our methodology is the following --
\begin{enumerate}
\item In order for a model to be useful over long time scales, we need to respect ``universal" physical laws. For instance, our model should be matter conserving and should not violate the laws of thermodynamics. Even tiny violations of these laws on short time scales can be amplified over time and lead to large errors or unphysical results over longer time scales. (Mass conservation is a non-trivial and basic requirement in oil reservoir simulations, particularly in chemically-reacting flows  \cite{zchen}). This is a particular concern for models that are entirely built from empirical rules obtained from measurements on short time scales.
\item As we argued above, we need to reduce the number of degrees of freedom to develop a model that can form the basis of a numerical simulation platform for oil spills.
\item While items (1) and (2) from above are certainly desirable, one has to pay a price to achieve them. This price is in terms of a tradeoff between computational feasibility and quantitative accuracy. In particular, empirical models with higher number of degrees of freedom might potentially run quicker and be more accurate for short times, than the mass conserving, reduced dimension model we develop in this paper. However, there are many situations where one needs to study the dynamics of oil spills over long periods of time, and it it then crucial that the model not give unphysical results.
\end{enumerate}

In its current incarnation, the oil transport model is a {\it shallow-water model}, {\em i.e.}, the horizontal spatio-temporal scales of the oil spill are much larger than the vertical scale. This does not imply that there are no significant vertical variations in distribution of the oil within the water column. Indeed, there are significant vertical variations due to the various processes, buoyancy, merging and splitting of oil droplets, turbulent dispersion, that act on the micro-scale. Our key contributions in this paper are that (i) we solve the relevant balance equations for the vertical distribution of oil in the water column; and (ii) we use the solutions to  {\em parameterize} the vertical distribution in terms of quantities (depth averages) that are slowly varying functions of the horizontal variables. These form the basis of an upscaling method that leads to effectively 2 dimensional modes for the oil transport with significantly fewer degrees of freedom.  

In order to couple this oil transport model with models of the ocean, we also demand that the oceanic flow be well captured by 2 dimensional models given by depth-averaging. This is not a significant restriction for practical application of our model. Indeed, familiar depth-averaged circulation models are successfully used to capture oceanic flows in estuaries, the continental shelf and shallow basins. These are regions of the ocean that abut fragile shore environments as well as large cities and population centers. These are also where large scale oil exploration takes place, as does complex shipping traffic. 

As we argue above, we view as a fundamental requirement that any practical  oil spill model with environmental relevance should conserve mass and describe the geographical distribution of oil over time. In essence should answer the most basic question of where and how much of the oil is to be found at any given time. It is impractical for existing observer network to capture how much oil 
is found because measurements sample the ocean interior and the surface very sparsely. Consequently, even in a data assimilation framework, it is crucial that the model conserve mass. The crucial ingredient in a mass conserving model is a proper accounting of the vertical exchange of oil between the surface and subsurface domains; this is the subject of this paper.

Although we develop our parameterization for the vertical distribution in the context of our shallow-water oil spill model, the parametrization itself or the ideas we use in the proposal could be ported to other oil models, even Lagrangian models.

This paper is organized as follows: We begin by briefly summarizing the shallow-water oil model and identify where the mass exchange term appears in the model. In RRV we focused on transverse dispersive and advective effects on oil spill dynamics. The mass exchange model, which is mentioned but not described in  RRV is proposed next.  The mass exchange model is formulated in terms of droplet size distributions in the bulk and it accounts for many distinct processes including buoyancy, vertical mixing due to wave breaking and boundary turbulence and merging/splitting of droplets. The size of a droplet plays a key role in its dynamics and thereby, in the oil model. Our focus is on large scale dynamics, meaning over spatio-temporal expanses that cover hundreds of kilometers and thousands of seconds. At first glance micro-mechanics in the oil could be perhaps ignored. However, surface tension effects, viscous effects, chemistry, and the background hydrodynamics leads to the breakup of oil into droplets.  
Because these droplets have a dynamic that is size dependent, it is not possible to ignore microscopic physics; for example, droplets that are very small take a very long time to buoyantly reach the sea surface. Whether droplets advect with the flow or are affected by shearing and compression also depend on the size of the droplets is accounted for in our model, as is merging and splitting in the subsurface oil.  The model of the process of droplet merging/splitting will describe how size distribution and droplet mechanics can be accounted for at the larger scales of the oil fate model.
In the Results  section  we revisit a nearshore phenomenon that we originally proposed by Restrepo {\it et al.}~\cite{sticky} (hereon, RVD) and have since used as a benchmark (see RRV) to test our refinements of the oil transport model. The {\it nearshore sticky-waters} phenomenon concerns the dynamics of oil approaching a nearshore environment and the conditions whereupon the oil slows down and even parks itself just beyond the break zone. The original model for nearshore sticky-waters~\cite{sticky} had a crude mass conserving rule for the exchanges of oil  mass exchanges between the slick and the sub-surface. The calculation shown here suggests that the original conceptual model  gives results that are consistent with the refined mass exchange model that we propose here. In the Summary section we   the mass exchange parametrization.

\section{The Shallow-Water Oil Fate Model}
\label{background}

%
%

The oil fate model itself is  described in RRV.  The general transport model is formulated for a surface slick layer $S$, in units of length,  and a subsurface component of oil, described by a concentration field $C$ in units of density times volume over volume. While there may be thousands of chemicals in a particular sample of oil and each oil spill may have different ratios of these chemicals, we describe the slick oil in terms of 
 \[
S = \sum_{i=1}^N s_i,
 \]
 where
$S({\bf x}, T)$, so that the total oil slick mass is given by 
 \[
 M_s(T) = \int_{\Omega_T} \sum_{i=1}^{N}  s_i({\bf x},T) \rho_i d {\bf x},
 \]
 ${\bf x}$ is the transverse coordinate,
 $\Omega_T$ is the (possibly time dependent) transverse domain. Time is denoted by $T$.
The  densities of each chemical complex are denoted by the  $\rho_i$. 

 For the $i^{th}$ chemical complex, the transport equation is 
\begin{equation}
\frac{\partial s_i}{\partial T} + \nabla_{\perp} \cdot \left( C_{xs} {\bf V} s_i + \frac{ {\bm \tau}}{2 \mu_i} s_i^2 \right) - \nabla_\perp \cdot [\Psi \nabla_\perp s_i] = R_i + E^s_i (s_i) 
+ P^s_i+ G^s_i.
\label{seq}
\end{equation}
Here, $\nabla_\perp$ is the transverse gradient operator, ${\bf V}$ are the $ {\bf x} $-components of the transport velocity, $C_{xs}$ is a slip parameter, 
${\bm \tau}$ is the wind stress, $\mu_i$ the viscosity of the $i^{th}$ oil species, $\Psi$ is the dispersion.
The last term on the left hand side is the eddy/dispersion term. 
 $R_i$ is the rate associated with  
mass exchanges between the oil slick and the  interior of the ocean  and is the focus of this study. 
The {\it reaction} term
$E^s_i(b_i,s_i,s_j)$ encompasses chemical reactions among chemical complexes within the slick as well as other processes modeled by rates, such as evaporation. 
 $P^s_i$ is the rate of biodegradation, emulsification, photodegradation 
 and $G^s_i$ is the source rate term.  Unlike $R_i$ the terms with  $s$ superscript   are particular to the surface oil.

The total mass of the oil in the sub-surface is defined as
\[
M_c = \int_{\Omega_T}  \int_{-h}^\eta \sum_{i=1}^N   c_i({\bf x},z,T) \, dz \, d {\bf x}.
\] 
where $ c_i $ is the concentration of the $ i $-th complex in the sub-surface layer, in units of mass per unit volume. Define the depth average concentration as
\begin{equation}
C_i({\bf x},T) = \frac{1}{H({\bf x})} \int_{-h}^{\eta(T)} c_i({\bf x},z,T) \, dz
\end{equation}
$H$ is the total water column depth.

The evolution of the $i^{th}$ chemical complex  in the subsurface is given by 
\begin{equation}
\frac{\partial {H C_i}}{\partial T} + \nabla_\perp \cdot (H {\bf V} C_i) 
-  \nabla_\perp \cdot [ H \Psi \nabla_\perp C_i]  = -\rho_i R_i +   \rho_i \left[E_i^C( C_i)   + P_i^C+ G_i^C\right].
\label{ceqfinal}
\end{equation}
On the right hand side, the first term is    the mass exchange term $ \rho_i R_i$, the reaction/chemistry term is $E_i^C$,  $P_i^C$ represents  the sedimentation and biodegradation mechanisms. The subsurface source term is $G_i^C$. 
 
 This paper concerns itself with developing a model for $R_i$, which we connote as the {\it mass exchange term} and has units of length per unit time. 
 To do so we do not need to carry around reference to chemical complexes, so we will just assume that there is a single chemical complex (we can then drop the subscript $i$ from $s$ and $C$).

\section{The Mass Exchange Model}
\label{model}

The mass exchange $R$ represents oil lost to the slick by turbulent mixing and wave breaking and oil gained from the subsurface by buoyancy. Because
buoyancy forces depend on the radius of the droplets, it is necessary to keep track of the droplet distribution in the subsurface, which in turn, requires that droplet merging and splitting be incorporated in the dynamics. The consequence of this is that micromechanics need to be resolved or they must be upscaled if the scales of interest are far larger than the droplet coherence scale. The oil model \cite{oil1} is being designed to capture the dynamics of oil spills at environmental  spatio-temporal scales, which are large. Hence, we will develop the rationale for an upscaled version of the mass exchange
process, applicable to environmental scales.


In summary, the mass exchange model takes into account mass exchanges between the slick oil and the subsurface oil as well as the dynamics of droplet merging and splitting in the subsurface. We will be making frequent reference to the study by Tkalich and Chan \cite{tkalichbreak} and thus we will refer to it as TC, hereon.

\subsection{Slick/Subsurface Oil Exchanges}
\label{slick}

Per unit area, at a fixed horizontal location $ {\bf x} $, the total mass of oil is
\begin{align}
M({\bf x},T) &= H({\bf x}) C({\bf x},T) + \rho s({\bf x},T) \\
&=  \int_{-h}^{\eta(T)} c({\bf x},z,T) \, dz +  \rho s({\bf x},T) \\
&=  \int_{-h}^{\eta(T)} \int_{0}^{\infty} \rho V(r) n(r,{\bf x},z,T) \, dr \, dz +  \rho s({\bf x},T) \\
&=  \int_{0}^{\infty} \rho V(r) \bar{n}(r,{\bf x},T) \, dr +  \rho s({\bf x},T)
\end{align}
where $ V(r) = \frac{4}{3}\pi r^3 $ and $ n(r,{\bf x},z,t) \, dr $ is the total number of droplets of radius in $ (r, r+dr) $ per unit volume of water at depth $ z $. 
$\bar{n}$ is the total number of droplets of radius in $ (r, r+dr) $ per unit area of water surface water, in the subsurface.
If we ignore the dispersion, reaction/chemistry, sedimentation, biodegradation and subsurface sources, the total derivative of $ M $ must be zero, and therefore
\begin{equation}
\frac{\partial HC}{\partial T} + {\bf V}\cdot \nabla(HC) + \rho R = -\rho \left[\frac{\partial s}{\partial T} + {\bf V} \cdot \nabla  s - R \right]
\end{equation} 
where
\begin{equation}
 R = - \int_{0}^{\infty} V(r) w_b(r) n(r,{\bf x},\eta,T) \, dr + \int_{-h}^{\eta} \int_{0}^{\infty} V(r) S_b \, dr \, dz.
 \end{equation}
 The first term of $R$ is the rate of oil gained by the slick (in units of length/time), coming from the subsurface by means of buoyancy. The second term is the
 rate of  oil lost (primarily) due to 
 wave breaking (see TC for an alternative  derivation). In the first term $w_b \ge 0$ is the buoyancy of oil, which is known to depend on the radius of the droplets. The droplet distribution is classified by $V(r)$ the droplet volume, which in turn depends on the radius $r$. The largest resolved radii is $r_0$ and the smallest is $r_m$. The second term (again, in units of length/time), is associated with the oil lost at the surface to the subsurface, with $S_b$ being 
 the {\it wave breaking source term}. 

\subsection{The wave breaking source term $ S_b $}

Wave breaking events occur at the location of interest at a rate $ \lambda_b $ and have a `strength' that may be characterized, for example, in terms of the slope $ h k $ of the breaking wave. For the time scales of interest, a wave breaking event occurring at time $ T_i $  can be regarded as an instantaneous transport event,  in which a fraction $ b_i $ of the volume $ s(T_i) $ is distributed into the water column. Let $ \delta_{T_i}(T) $ denote the delta function at $ T_i $, $ \delta_{T_i}(T) = \frac{1}{dT} $ for $ T \in [T_i, T_i + dT) $, and zero everywhere else. The breaking inputs can then be written in terms of a compound Poisson $ N_b $ process of rate $ \lambda_b $,
\begin{equation}\label{Def_b}
b(T) = \sum_{i=1}^{N_b(T)} b_i \delta_{T_i}(T)
\end{equation}
We will refer to $ b_i $ as the \textit{strength} of the $ i $th wave breaking event.

According to Delvigne and Sweeney \cite{delvignesweeny88}, if a unit of oil per unit area is available to be distributed into droplets, the number of droplets with radius in $ (r,r + \,dr) $ will be $ N(r) \,d r $ with
 \begin{equation}\label{Def_Nr}
  N(r) :=\frac{3(4-s)}{4 \pi (r_m^{4-s} - r_1^{4-s})} r^{-s}, \; r \in [r_1, r_m],
  \end{equation}
zero otherwise, which is such that $ \int_{0}^{\infty} V(r) N(r) \,d r = 1 $.

$ S_b(T,r,z) \,dz $ is the rate, per unit time, at which droplets of radius in $ (r,r+\,d r) $ become available in the volume $ (z,z+\,dz)$ at time $ T $. This term is nonzero only at breaking times $ \{T_i:i=1,2,\dots\} $. At $ T_i $, the unit of ocean surface area we are considering suffers a breaking event displacing a volume $ b_i s(T_i) $ of oil to be distributed along the depth of water. Assume that such distribution is independent of everything else and is given by $ g_b(z) $ with 
\begin{equation}\label{Eqn_Intg}
\int_0^\infty g_b(z) \,d z =1. 
\end{equation}
The volume of oil available in the volume $ (z,z+\,d z)$ of water is $ b_i s(T_i) g_b(z) \,dz $, and by \eqref{Def_Nr}, the resulting rate of available droplets of radius in $ (r, r+\,d r) $ is 
\begin{equation} \label{Def_Sb}
 S_b(T_i,z,r) = \frac{1}{dT} N(r) g_b(z) b_i s(T_i). 
\end{equation}

By \eqref{Eqn_Intg} and \eqref{Def_Nr} we can write the mass exchange term as
\begin{equation}
R({\bf x},T) = - \int_{0}^{\infty} V(r) w_b(r) n(r,{\bf x},\eta,T) \, dr + \sum_{i=1}^{N_b(T)} s(T_i) b_i \delta_{T_i}(T). 
\label{req}
\end{equation}

\subsection{Subsurface Droplet Distribution Dynamics}
\label{droplet}

We must now specify a dynamical model for the droplets in the subsurface. In accordance with the oil model~\cite{oil1}, droplets are assumed to undergo a linear vertical advection-diffusion process. Furthermore, the time scales at which the vertical distribution of oil sorts itself out, given fast disturbances at the sea surface, are assumed to be very short compared to the horizontal advection scales (the shallow water wave approximation).

We denote the turbulent diffusivity by $ D(r)  \sim 10^{-2} - 10^{-3} \SI{}{\meter\per\second\squared}$ which can be parametrized in terms of the turbulent description of the flow, while $ w(r) \sim 10^{-6}-10^{-3} \SI{}{\meter \per \second} $ is the limiting velocity of a oil droplet moving through water as described by TC (Equation 13).

 In what follows we focus on vertical balances, an adiabatic approximation, in which for given radius $ r $, depth $ z $ and time $ T $, two flux sources  affect $ n(r,{\bf x},z,t) $: the droplets that are entrapped from the surface into depth $ z $ after a wave breaking event, and changes on the distribution of radii due to merging and splitting of droplets. Putting all together, gives
\begin{equation}\label{Eq_n}
\frac{\partial n}{\partial T} =  D_z(r) \frac{\partial^2 n}{\partial z^2} - (W({\bf x},z,T) + w_b(r)) \frac{\partial n}{\partial z} + S_b + \mathcal{K}[n] ,
\end{equation}
Furthermore,  $W \ll w_b(r)$, hence we will ignore $W$. (We showed in RRV  that depth integration effectively dispenses with $W$).
The operator $ \mathcal{K} $ encodes the effect of merging and splitting of droplets. This operator is discussed in detail in section \ref{breakup}, but for now, it suffices to remark that it should be conservative, namely:
\begin{equation}\label{Eq_hcons}
\int_{0}^{\infty} \mathcal{K}[n](r)\,dr = 0.
\end{equation}

At the surface $ z=\eta(T) $ we assume that there is no diffusive flux as the slick is saturated, while all incoming droplets from the upward effect of buoyancy are added to the volume $ s $ and removed from the concentration $ n $. The boundary condition at the subsurface/slick interface is simply:
\begin{equation}\label{Eq_bc_cw}
\frac{\partial n}{\partial z}(t,\eta,r) = 0, \quad t>0, \; r \in [0,\infty). 
\end{equation}
At $z=-h$ we allow for  sedimentation, which we can ignore for now.

In what follows we fix $ {\bf x} $ (which we omit in the subsequent notation), and 
solve for the amount of oil in the slick and subsurface droplets occupying the one-dimensional moving water column $ (-h, \eta(T)) $, ignoring as well horizontal advection/dispersion. Specifically, we solve
\begin{align}
\frac{\partial n}{\partial T} &=  D_z(r) \frac{\partial^2 n}{\partial z^2} - w_b(r) \frac{\partial n}{\partial z} + \sum_{i=1}^{\infty} s(T_i) b_i \delta_{T_i}(T) + \mathcal{K}[n] \label{Eq_n_z}  \\
\frac{\partial s}{\partial T} &=  \int_{0}^{\infty} V(r) w_b(r) n(r,{\bf x},\eta,T) \, dr - \sum_{i=1}^{\infty} s(T_i) b_i \delta_{T_i}(T) \label{Eq_s}.
\end{align}

The solution to \eqref{Eq_n_z} is:
\begin{eqnarray}
n(r,z,T) &=&  \sum_{i=0}^{N_b(T)}  \int_{-h}^{\eta(T_i)} g_b(z') G(r,z',z,T-T_i) \,d z' \, N(r) s(T_i) b_i \nonumber\\
&+ &\int_0^T \int_{-h}^{\eta(T')} G(r,z',z,T-T') \mathcal{K}[n(\cdot,z',T')](r) \,d z' \,d T' \label{Eq_Sol_n}
\end{eqnarray}
where we have made the convenient choice for the initial condition 
\begin{equation}
n(0,z,r) = \frac{1}{\rho}H C(0) N(r) g_b(z)
\end{equation}
and set $ T_0=0,\, b_0 = \frac{1}{\rho} \frac{HC(0)}{s(0)} $. The function $ G $ in \eqref{Eq_Sol_n} is the Green's function associated with the vertical terms of the parabolic equation \eqref{Eq_n} of $ n $ which may be approximated with the Greens' function corresponding to the operator $ w_b \partial_z + D_z \partial^2_z$ on $ (-\infty, \eta) $, with boundary condition \eqref{Eq_bc_cw}. That is to say,
\begin{align}
G(r,z,z',T) =& \frac{1}{2  \sqrt{\pi T
   D_z}}\left\{\exp\left[-\frac{(z'-z)^2}{4 T
   D_z}\right]+\exp\left[-\frac{(2 \eta-(z+z'))^2}{4 T
   D_z}\right]\right\} \nonumber\\
   &\times \exp\left[-\frac{w_b
   (2 (z'-z) +T w_b )}{4
   D_z}\right] \nonumber\\
   &-\frac{1}{2 D_z} w_b \exp\left[\frac{(\eta-z') 
   w_b}{D_z}\right]
   \text{erfc}\left[\frac{2 \eta-(z+z')+T w_b}{2
   \sqrt{T D_z}}\right].
\end{align}

In the simplest of possible situations, we illustrate the dynamics of the mass exchange and its impact on the slick and subsurface oil. For simplicity we assume $ \mathcal{K}\equiv 0 $ and propose a simple backwards difference scheme for  he approximate solution of $ s $, via  (\ref{Eq_s}):
\begin{align*}
s(T) &\approx  \frac{1}{1+\Delta T \, b_i \delta_{T_i}(T)} (s(T-\Delta T) + \Delta T \, F_0(T)), \quad \mbox{where}\\
F_0(T) &= \sum_{i=0}^{N_b(T)}   s(T_i)  \, b_i \delta_{T_i}(T) \int_{r_1}^{r_m} V(r) w_b(r) N(r)  \int_0^{\infty} g_b(z') G(T-T_i,z',0,r) \, d z' \, d r.
\end{align*}
Here we use the parametrization of $ w(r)$ and $ D(r) $, suggested by TC, 
$w(r) = \frac{2g(1-\rho/\rho_w)}{9 \nu_w}r^2$, 
$D(r) = D_0 \left(\frac{D_0}{D_m} \right)^{-\frac{r}{r_m}}$, 
with  $ \rho =\SI{800}{\kilogram \per \meter \cubed}$, $D_0 = 10^{-2}\SI{}{\meter \per \second \square}$, 
$D_m = 10^{-3}\SI{}{\meter \per \second \square}$, $ r_1 = 10^{-6}\SI{}{\meter} $, $ r_m = 10^{-4}\SI{}{\meter} $.
We picked the rate of breaking as  $ \lambda_b = 10^{-3}\SI{}{\per \second} $ and the fraction of entrapped oil randomly distributed as $ b_i \sim \text{Uniform}(0, 10^{-1}). $ After a breaking event, the entrapped volume is distributed exponentially over the depth so 95\% of it is above $ z_m = \SI{1}{\meter}$, $ g_b(z) = \kappa e^{-\kappa z}$ with $\kappa = \SI{3}{\per \meter}. $

\begin{figure}[bht]
\centering
\includegraphics[scale=1.25]{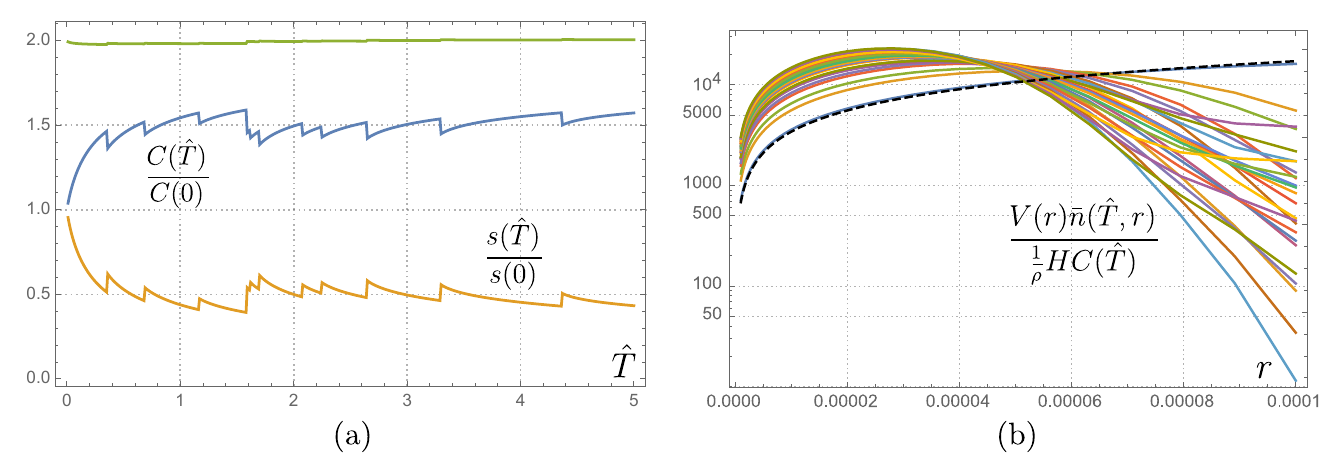}
\caption{Evolution of non-dimensional surface and subsurface oil. (a) $ s/s(0) $ in blue, $ C/C(0) $ in yellow, and their sum in green. The sharp changes coincide with the breaking events. Mass is conserved (green), to within acceptable numerical accuracy. Non-dimensional quantities are $ \hat{T} = \frac{D_m}{z_m^2} T $,  $\hat{w_b} = \frac{z_m }{D_m}w_b $,  $\hat{D} = \frac{D}{D_m}$,  $\hat{\lambda}_b = \frac{z_m^2}{D_m} \lambda_b$. (b) Fraction of volume of oil entrapped as a function of radius for selected times, see equation \eqref{Def_nbar} for $ \bar{n} $. The initial density is shown as dashed. The values of large $ r $ go down in between breaking events and kick up at times of breaking. The fraction of small droplets steadily increases with time. }\label{Fi_CsCw}
\end{figure}

For this illustrative example, the evolution of the fraction of slick and subsurface oil, as well as the distribution of the radii in the subsurface is shown in Figure \ref{Fi_CsCw}. There is an increase in the droplet number for small radii. 

\section{Depth-Averaged Dynamics}

The total amount of submerged particles of radius in $ (r,r+dr) $ is
\begin{equation}\label{Def_nbar}
\bar{n}(r,T) = \int_{-h}^{\eta(T)} n(r,z,T) \, dz,
\end{equation}
and is given by
\begin{eqnarray}
\bar{n}(r,T) &=&  \sum_{i=0}^{N_b(T)}  \int_{-h}^{\eta(T_i)} g_b(z') \bar{G}(r,z',T-T_i) \,d z' \, N(r) s(T_i) b_i  \nonumber\\
&+ &\int_0^T \int_{-h}^{\eta(T)} \int_{-h}^{\eta(T')} G(r,z',z ,T-T') \, \mathcal{K}[n(\cdot, z',T')](r) \,d z' \, dz \,d T.' \label{Eq_Sol_nbar}
\end{eqnarray}
Here,
\begin{equation}
\bar{G}(r,z',T) = \int_{-h}^{\eta(T)} G(r,z',z,T) \, dz.
\end{equation}

We would like to derive an expression for the dynamics of merging and splitting of droplets that is independent of $ z $ and hence arrive to an approximate version of \eqref{Eq_Sol_nbar} in terms of solely $ \bar{n} $. We may assume that the merging and breaking processes occur at time scales much shorter than those of $ T $. In particular, during an interval $ dT' $, the change in the distribution of radii is
\begin{equation}
n(r,z',T' + dT') - n(r,z',T') \approx \mathcal{K}[n(\cdot,z',T')] dT'.
\end{equation} 
Out of these added droplets, the following amount is then transported to depth $ z $
\begin{equation}
\int_{-h}^{\eta(T')} G(r,z',z ,T-T') \, \mathcal{K}[n(\cdot,z',T')](r) dT' \,d z'.
\end{equation}
We approximate the second integral in \eqref{Eq_Sol_nbar} by conmuting these mechanism, and writing
\begin{align*}
&\int_0^T \int_{-h}^{\eta(T)} \int_{-h}^{\eta(T')} G(r,z',z ,T-T') \, \mathcal{K}[n(\cdot,z',T')](r) \,d z' \, dz \,d T'\\
&\approx \int_0^T \int_{-h}^{\eta(T)} \mathcal{K} \left[ \int_{-h}^{\eta(T')} G(\cdot,z',z ,T-T') \, n(\cdot,z',T') \,d z'  \right] \, dz \,d T' \\
&= \int_0^T  \mathcal{K} \left[ \int_{-h}^{\eta(T)} \int_{-h}^{\eta(T')} G(\cdot,z',z ,T-T') \, n(\cdot,z',T') \,d z' \, dz \right]  \,d T' \\
&\approx \int_0^T  \mathcal{K}[\bar{n}(\cdot, T')]  \,d T',
\end{align*}
which essentially disregards the depth-wise structure of $ n $ when it comes to the merging and splitting mechanism. The equation for $ \bar{n} $ is now
\begin{equation}
\bar{n}(r,T) =  \sum_{i=0}^{N_b(T)}  \int_{-h}^{\eta(T_i)} g_b(z') \bar{G}(r,z',T-T_i) \,d z' \, N(r) s(T_i) B_i + \int_0^T  \mathcal{K}[\bar{n}(\cdot, T')](r)  \,d T' .\label{Eq_Sol_nbar2}
\end{equation}

When the oil spill is young, or the sea state is rough, droplet distributions change quickly due to merging and splitting.  Further, 
it is well known that droplets in the subsurface smaller than a critical radius, will remain in the subsurface  (in TC the critical radius for oil is quoted as $r_c=50$ $\mu$m).
For computational reasons it is convenient to bin droplet sizes for each chemical complex and formulate the subsurface oil transport in terms of 
a finite, discrete set of droplet classes. (Tracking the oil slick droplet sizes is unnecessary). At the same time, we will incorporate depth-averaging
and pose the dynamics of subsurface droplets in terms of fluxes into and out of the depth-averaged horizontal computational cells 
due to advection/dispersion, interactions with the slick via buoyancy and wave mixing, and finally via changes to droplet class distributions within each cell.

 Let $ [r_j, r_{j+1})$, $ j =1,\dots, N_r :=m-1$ denote different radius ranges to be considered and denote:
\begin{equation}
\bar{n}_j({\bf x}, T) = \int_{-h}^{\eta(T)} \int_{r_j}^{r_{j+1}} n(r,{\bf x},z,t) \, dr \, dz.
\end{equation}
Ignoring advection and dispersion, 
an equation for the evolution of $ \bar{n}_j $ can be obtained via equation \eqref{Eq_n}:
\begin{eqnarray}
\frac{\partial \bar{n}_j}{\partial T} &=& - \int_{r_j}^{r_{j+1}} w_b(r) n(r,{\bf x},\eta,T) \, dr + \sum_{i=1}^{N_b(T)} s(T_i) B_i \delta_{T_i}(T) \int_{r_j}^{r_{j+1}} N(r) \, dr  \nonumber \\
&&+  \int_{-h}^{\eta} \int_{r_j}^{r_{j+1}} \mathcal{K}[n] \, dr \, dz
\end{eqnarray}

We now make the following approximations: we use $ w_b(r) \approx w_b(r_j) $ for all $ r \in [r_j,r_{j+1}) $. Lastly, we assume that at the time scale of interest, we can write
\begin{equation}
n(r,{\bf x},\eta,T) \approx \frac{1}{H} \int_{-h}^{\eta} n(r,{\bf x},z,T) \, dz.
\end{equation}
The operator $ \mathcal{K} $ can be simplified, by ignoring depth dependence and we can thus propose a dynamic between size classes
that takes into account droplet merging/splitting and interactions with the eddies in the flow, within the subsurface compartment. We then arrive to
\begin{equation}
\frac{\partial \bar{n}_j}{\partial T} = - \frac{w_b(r_j)}{H} \bar{n}_j + \sum_{i=1}^{N_b(T)} N_j s(T_i) b_i \delta_{T_i}(T) +  \frac{1}{H} B_j - \frac{1}{H} D_j,
\label{mastern}
\end{equation}
where $ N_j := \int_{r_j}^{r_{j+1}} N(r) \, dr $, and $j=1,2..,m$. And similarly for the slick:
\begin{equation}
\frac{\partial s}{\partial T} = \sum_{j} \frac{V(r_j) w_b(r_j)}{H} \bar{n}_j - \sum_{i=1}^{N_b(T)} s(T_i) b_i \delta_{T_i}(T).
\label{shalfstep}
\end{equation}
The terms $B_j$ and $D_j$ are birth and death rates for droplets in class size $j$. These two terms will be developed fully in Section \ref{breakup}.

\subsection{Subsurface Oil Droplet Merging and Coalescence}
\label{breakup}

We will focus now on the merging/splitting mechanism and ignore all other effects. Let us consider the situation for a subsurface cell
at position  ${\bf x}$. The changes to the 
total number of droplets  of size class  in $ (V(r),V(r+dr)) $ occupying a fixed volume of water, due to breaking and merging obey the balance equation
\begin{equation}
\frac{\partial \bar n(V,T)}{\partial T} = (B_V - D_V)  =:  \mathcal{K}[\bar n] ,
\label{eq:population_blance_eq}
\end{equation}
where  $D_V, B_V$ are death and birth rates, collectively defined as  $\mathcal{K}[\bar n]$, that approximate the Smoluchowsky operator $ \mathcal{K}[n] $. The death $D_V$ and birth $B_V$ terms, in turn, are defined as 
\begin{align}
& \hspace{-1cm} D_V = \bar n(V,T) \int_0^\infty \lambda(V,V')\, h(V,V') \bar n(V',T)dV' +g(V)\, \bar n(V,T) \label{eq:death} \\
\begin{split}
B_V = &\int_0^{\frac{V}{2}} \lambda(V-V',V')\, h(V-V',V') \bar n(V-V',T) \bar n(V',T)dV' \\
     &+ \int_V^\infty \beta(V,V')\, \nu(V') \, g(V')\, \bar n(V',T) dV'
\end{split}
\label{eq:birth}
\end{align}
where $V:=V(r)$, $V':=V(r')$. In the above expressions $\lambda(V,V')$ is the coalescence efficiency for collisions between droplets of classes $V$ and $V'$, $h(V,V')$ is the frequency of droplets of class $V$ 
colliding with droplets of class  $V'$, 
$g(V')$ is the breakage frequency of droplets of class $V'$,
$\nu(V')$ represents the number of droplets formed by breakage of droplets of class $V'$ and $\beta(V,V')$ is the probability of forming droplets of class $V$ from breakage of droplets of class  $V'$. To first order, it can be assumed  that only binary collisions occur, and thus $\nu(V') = 2$.

Once the droplet distribution is binned (assuming the total number of bins $m$ is even), 
and the distribution of droplet sizes $r_1$ and $r_m$, is labelled by $j=1,..,m$.  The $H D_j(\cdot, T) \approx D_{V(r_j)}$ and 
$H B_j(\cdot,T) \approx B_{V(r_j)}$ are:
\begin{eqnarray}
H D_j &=& \bar n( r_j,T) \sum_{l=1}^{m} \lambda(r_j,r_l) h(r_j,r_l) \bar n(r_l,T) + g(r_j) \bar n(r_j,T) \label{Heq} \\
H B_j &=& \sum_{l=1}^{m/2} \lambda (r_j-r_l,r_l) h(r_j-r_l,r_l) \bar n( r_j-r_l,T) \bar n( r_l,T)  \nonumber \\
&+& \sum_{l=j}^m \beta(r_j,r_l)g(r_l) \bar n( r_l,T). \label{Beq}
\end{eqnarray}
\subsubsection{Parametrizing Breakup Mechanics}

Droplet breakup is caused when an  external force in the surrounding fluid exceeds the  surface and internal forces in the droplet \citep{liao2009literature}. 
Specific breakup  mechanisms are  droplet  elongation in sheared flow, turbulent pressure fluctuations, relative velocity fluctuations,  and droplet-eddy collisions. The breakup frequency, $f(d)$, assuming droplet-eddy collisions in isotropic turbulence conditions, droplet sizes in the inertial sub-range,  and droplet breakup due solely by eddy-droplet collisions with  smaller or the comparably-sized eddies is defined as:
\[
f(r) = K_b \, h(r)\, B_e .
\label{eq:breakage_freq}
\]
 $K_b$ is a system dependent parameter determined as calibration factor \citep{zhao2014vdrop}.

The eddy-droplet collision frequency, $h$, is defined as:
\begin{equation}
h(r)=\int_{n_{e,m}} S_{ed} (u_e^2 + u_r^2)^{\frac{1}{2}} \bar{n} \, d n_e,
\label{eq:eddy-drop-coll_freq}
\end{equation}
where $S_{ed} = \pi (r_e+r)^2$ is the collision cross-section area, $r_e$ is the eddy size which collide with a droplet of radius $r$, $dn_e$ is number of eddies of size between $r_e$ and $r_e+ d  r_e$. 
 Here,  $u_e^2 = 3.8 (\epsilon / k)^{\frac{2}{3}}$ is an estimate of the  velocity-squared of the eddy, $k   = \pi/r_e$ is wavenumber, 
and $u_r^2 = 1.7 (\epsilon r)^\frac{2}{3}$ is the droplet velocity.
The estimated number of eddies per unit volume of the fluid is  found from $dn_{e,m}(k)=3\times 10^{-3} k^2 dk$  (see \citep{azbel1981two}).

The breakup efficiency $B_e$ is the probability of breakup if a drop-eddy collision takes place:
\[
B_e=\exp \left[- \frac{E_c + E_v}{c_1e} \right] .
\label{eq:breakage_eff}
\]
The minimum energy for droplet breakup is related to the amount of required energy to keep the surface of a droplet intact. 
For binary breakage, this energy is assumed  equal to the mean energy required to break the drop into two equally-sized droplets (maximum energy) and into a one small and  one big droplets (minimum energy) \citep{tsouris1994breakage}. Hence,
\[
E_c=2  \left[ 2\pi \sigma_d \left( \frac{r}{2^{1/3}} \right)^2+ \pi \sigma_d r^2_{\max}+ \pi \sigma_d r^2_{\min} -2 \pi \sigma_d r^2  \right] 
\label{eq:surface_tension_min_energy}
\]
where $e = \frac{1}{2}mu_e^2 = \frac{2}{3}\pi \rho_w r_e^3 u_e^2$, $r_{\min}=r_1$, $r_{\max} = r_m$, and $\sigma_d $ is the droplet surface tension
 \citep{zhao2014vdrop}.

Oil droplets are subjected to surface tension effects as well as internal viscous forces. Hence we need to take into account the balance of these
(see  \citep{calabrese1986drop}). 
The energy associated with this internal force is
\begin{equation}
E_v= \left[1+a \exp \left(-\frac{T}{T_{tot}} \right) \right] \left[ \frac{\pi}{6} \epsilon^{\frac{1}{3}}(2r)^{\frac{7}{3}} \mu_d \sqrt{\frac{\rho_w}{\rho}}   \right],  
\label{eq:internal_viscos_energy}
\end{equation}
where $\rho_w$ is the fluid density, $\rho$ is the density of  the particular chemical complex oil  density,  $T_{tot}$  is interaction simulation time (one time step of fluid-oil interaction), and $\mu_d$ is the oil species viscosity. The constant $a = 1.1$ is a calibration constant (see Zhao {\it et al.}~\cite{zhao2014vdrop}, Equation 15).


In Tsouris and Tavlarides~\citep{tsouris1994breakage} it is argued that, assuming binary breakage (recall it was  assumed that  $\nu(V')$ is 2), the amount of energy required for the generation of two like-sized droplets is larger  than what is required to produce one small and one big droplets. They suggested the breakup PDF  of a drop  of class size $V$,  from a parent drop of class size $V'$ is given by:
\begin{equation}
\beta(r,r')=\frac{E_{\min} + \left[ E_{\max}-E(r) \right] }   {\int_0^{r'} \left\{ E_{\min} + \left[ E_{\max}-E(r) \right] \right\} \, d r },
\label{eq:breakage_probability_density}
\end{equation}
where: 
\[
E(r)=4 \pi \sigma [(r'^3 -r^3)^{2/3}+r^2-r'^2].
\label{eq:breakage_energy}
\]

\subsubsection{Parametrizing  Coalescence Mechanics}

The merging process in which two (or more) colliding droplets form a bigger droplet is called coalescence. The coalescence of droplets normally evolves in three stages: (1) collision; (2) interfacial drainage of liquid at touching  interfaces;  and (3) the rupture of  the droplet  surface.  
In this study  we assume that background turbulence is the main reason for the collisions. 
The volume absorbed  by a moving droplet when  colliding with other droplets is the basis for calculating the collision frequency. 
The swept volume rate $h(V,V')$ is modeled following the kinetic gas theory. Assuming both droplets moving with turbulent eddies with relative velocity of $c$, a collision tube with cross section of $\pi(r+r')^2$ is swept. Therefore, the $h(V,V')$ could be defined as:
\[
h(V,V'):=\pi(r+r')^2 c.
\label{eq:col_freq}
\]
Assuming 
 isotropic turbulence   \cite{Kolmogorov41},  in Kuboi {\it et al.}~\cite{kuboi1972behavior} the relative colliding velocity  in stirred and pipe flows was  estimated as:
\[
c^2=2  \epsilon^{2/3} (r+r')^{2/3}.
\label{eq:col_vel}
\]
Hence the swept volume rate for droplets $V$ and $V'$ reads:
\[
h(V,V')=\frac{\pi}{4} 5.657 \epsilon ^{1/3} (r+r')^{7/3}.
\label{eq:col_freq2}
\]
The droplets need to stay in contact for a certain time to complete the coalescence process. Turbulence promotes  collisions, however, it can also 
move the droplets  away from each other before the coalescence process takes place. Therefore the collision efficiency is defined as the ratio between the average drainage time $\bar{t}$ and average contact time $\bar{\tau}$:
\[
\lambda(V,V')=\exp \left\{-\frac{\bar{t}}{\bar{\tau}} \right\}.
\label{eq:col_eff}
\]
(See Prince and Blanch~\citep{prince1990bubble}).
The drainage time of the thin interface between two droplets depends on the viscosities of each droplet and the main phase (could be different for air bubbles in comparison with oil droplets).  As a result we need to be more careful about implementation of this parameter.
The average drainage time for fully mobile interfaces is given by \cite{prince1990bubble}, Equation 18 :
\[
\bar{t}= \left(  \frac{R_{eq}^{3} \, \rho_l}{16 \sigma_d} \right) ^{1/2} \ln \frac{h_0}{h_c},
\label{eq:drainage_time}
\]
where $\rho_l$ is the density of the continuous phase, $\sigma_d$ is the surface tension and $h_0$ is the initial film thickness. The critical film thickness 
 $h_c$, as estimated  in 
 \citep{chesters1991modelling}, is  
\[
h_c= \left( \frac{A \, R_{eq}}{8\pi\sigma_d}\right ),
\label{eq:tick_critic}
\]
where $A$ is the Hamaker constant and $R_{eq}$ is the equivalent droplet radius, given by:
\[
R_{eq}=\frac{2}{1/r + 1/r'},
\label{eq:radius}
\]
where $r$ and $r'$ are the radii of the colliding droplets. 
The average contact time $\bar{\tau}$  is given by
\[
\bar{\tau}= 2^{2/3} \frac{(r+r')^{2/3}}{\epsilon^{1/3}},
\label{eq:contact_time}
\]
(see  Levich~\cite{levich1962physicochemical}).

\subsection{The Complete Depth-Averaged Model}
\label{themodel}

We now summarize the physical/computational shallow-water oil fate model. We use a split-step strategy, with time step $\Delta T$. 
\begin{itemize}
\item  For each chemical complex, at some time $T+ \Delta T/2$, 
 the slick is updated and denoted by   $s^*$. This is  done using $s(\cdot,T)$  and   (\ref{shalfstep}) (the sum runs for $j=1,\ldots,m$).
 For the subsurface, we produce the update $\bar n^*_j$ via (\ref{mastern}) knowing $\bar n_j(\cdot,T) $.
\item To step to $T+\Delta T$, we proceed as follows: Using $s^*$ we update $s(\cdot, T+\Delta T)$ 
using (\ref{seq}) with no $R$ (it has already been taken
into account in the half step). We also update  the subsurface concentration: we replace 
$H C(\cdot, T+\Delta T) = \sum_{j=1}^m \rho V_j \bar n_j(\cdot, T+\Delta T) \Delta r$ in (\ref{ceqfinal}), with $R=0$, and use
 $C^*:= \rho \sum_{j=1}^m V_j  \bar n^*_j \Delta r$ with $\Delta r = R_r/m$, where $R_r$ is the range of radii of the oil droplets, and $m$ is the
 number of bins used in partitioning the distribution of oil droplets, classified by their radii.  
\end{itemize}

In total, for each chemical complex there will be 1 evolution equation for the slick, $m$ for the subsurface. All told $N\times(m+1)$ degrees of freedom, where $N$ is the number of chemical complexes.

\section{Results}

We first benchmark the mechanics of merging and coalescence in our model by comparing its predictions to the results shown in
Tsouris and Tavlarides~\citep{tsouris1994breakage}, since this work largely inspired the droplet merging and breaking  dynamics of our oil model.
We then explore the dynamics within a single stack. A stack is composed of a surface compartment, and a subsurface compartment, directly below. The surface compartment corresponds to the slick. The subsurface compartment has oil and water. In the stack computations that follow, we  will feature the birth/death dynamics as well as the mechanics between the slick and the subsurface compartments. Since it is a single stack, no advection and dispersion takes place. 

Finally, we explore  a simple nearshore ocean/oil dynamics problem related to {\it nearshore sticky waters}. In RVD we proposed a conceptual model for the slowing down and possible  parking of buoyant material, such as oil, as it approaches the nearshore. The model suggests that oil being advected by ocean currents 
will be largely subjected to dispersive effects that gain prominence near the shore. We specifically claim that the dynamics of nearshore oil cannot be properly captured if dispersion is ignored. In fact, it is essential in explaining how oil reaches the shore even when the depth-averaged velocity normal to the shore goes to zero, if the oil cannot establish a conveyor dynamic near the shore. We have been replacing simple conceptual components of the original model proposed in RVD  to show that, (1) as the mechanics of the model are refined,  the proposed dynamics for sticky waters still holds; (2),  how the oil model proposed in RRV captures the phenomenon of nearshore sticky waters. In the calculations that will follow the stack 
benchmark we will highlight how dispersion, advection, and now droplet dynamics play out in the nearshore context; (3) we also want  to suggest that the nearshore sticky water problem, which is in itself an important dynamics problem for pollution transport close to the shore,  could be used as a field problem to test many aspects of oil transport dynamics and models.

\subsection{Benchmarking the Merging and Coalescence Mechanics in our Model}
\label{benchmark}

\begin{figure}[bht]
\centering
\includegraphics[scale=1.2]{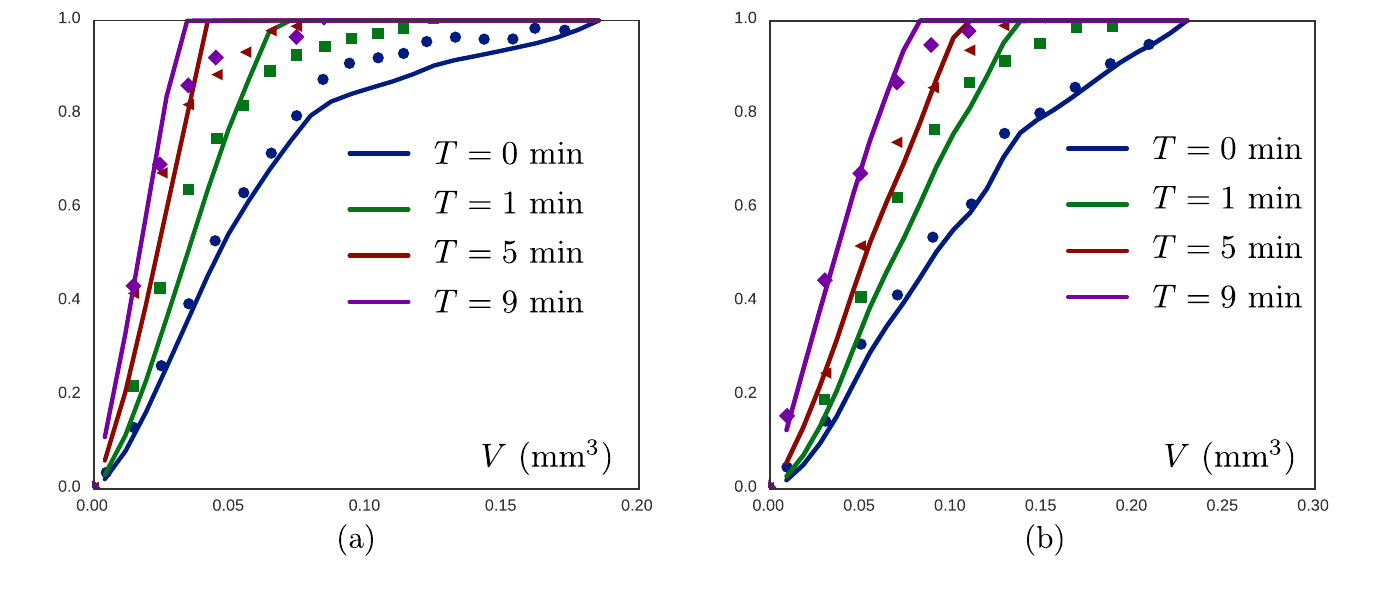}
\caption{Comparison of the time evolution of the cumulative droplet distribution $ \int_{V_1}^{V} \bar{n}(V',T) d V' $ of data and of predictions  by  Equation (\ref{eq:population_blance_eq}). Experimental data was taken from   \citep{tsouris1994breakage} and reported originally in \cite{baetavlarides}.
(a) Comparison of our model results with Figure 3 and Figure 4 of \citep{tsouris1994breakage}.  (a)  Corresponds to  a ratio of oil-to-water of 0.1; (b) corresponds to an oil-to-water ratio of 0.2. }
\label{t34}
\end{figure}

We benchmarked the birth/death mechanism in our model by comparing our results to those of 
in Tsouris and Tavlarides~\citep{tsouris1994breakage}. Rather than replicating here what Tsouris and Tavlarides did in the comparison of their model to data, we
suggest the reader refer to  Tsouris and Tavlarides~\citep{tsouris1994breakage} for the details.  In order to accomplish a similar  comparison between our model and data we omit the slick dynamics, and suppressed all of the mechanics in the subsurface dynamics, except for 
the terms associated with $B$ and $D$. Hence, we are using only (\ref{eq:population_blance_eq}) in this benchmark comparison. In the calculations we adopted the same parameter values as those used to generate Figures 3 and 4 in  Tsouris and Tavlarides~\citep{tsouris1994breakage}:
For continuous phase (water):  the viscosity is  $\mu_c = 1.787 \times 10^{-3}$ Ps,  water density $\rho_{w}$  = 1025.0 kg/m$^3$, and the
kinematic viscosity $\nu_c$  = $\mu_c / \rho_w$  m$^2$/s. For  the dispersed phase (Toluene), we choose a
surface tension $\sigma_d$ = 28.4 $\times 10^{-3}$ N/m,    
a dispersed phase molecular viscosity $\mu_d$     = 0.59 $\times 10^{-3}$   Ps , and a density  $\rho$  = 865    kg/m$^3$.
The droplets ranged in size from  $r_{1}$ = 0.0001 m to  $r_{m}$ = 0.00038, and we used $ m-1 = 25$ size classes.
 In Figure \ref{t34}  we superimpose the output of our model (lines) with digitized disconnected dots representing the experimental data used in
Tsouris and Tavlarides~\citep{tsouris1994breakage} and originally obtained by Bae and Tavlarides~\cite{baetavlarides}.
Our results show similar consistency between model and data as was reported by Tsouris and Tavlarides, when they compared their model to the same data, especially when we compare the evolution of the smaller size classes. We note that our model is considerably simpler than theirs, and apparently the simplification degrades the performance of our model for the larger size classes. This degradation of  fidelity is acceptable and deliberate: for the large spatio-temporal scales demanded by the applications of our oil fate model, gains in computational efficiency at the expense of small degradations in fidelity is an acceptable tradeoff. 
 
\subsection{Mass Exchanges in a Single Stack}
\label{column}

We consider next the dynamics of a stack, under equations (\ref{mastern}) and (\ref{shalfstep}).
For this case the size classes were chosen using $r_{1}$ = 0.0001 m and  $r_{m}$ = 0.00038 m   and a  volume ratio of $V_i/V_{i-1}$ = 1.5 resulting in 9 different size classes. We simulated one hour of random wave breaking given by \eqref{Def_b}  with $\lambda_b = \frac{1}{6T_w}$ ($ T_w $ being the wave period, see Restrepo {\it et al.}~\cite{RMMB}). 
The material properties of the oil and the water are the same as was used in the preceding benchmark calculation.  We also initialized oil at the surface and inside the mixing layer with equal
volume to follow the dynamical interaction between surface and sub-surface oil wave breaking strength $ b_i $ of varying ranges.

\begin{figure}[hbt]
 \centering
\includegraphics[scale=1.15]{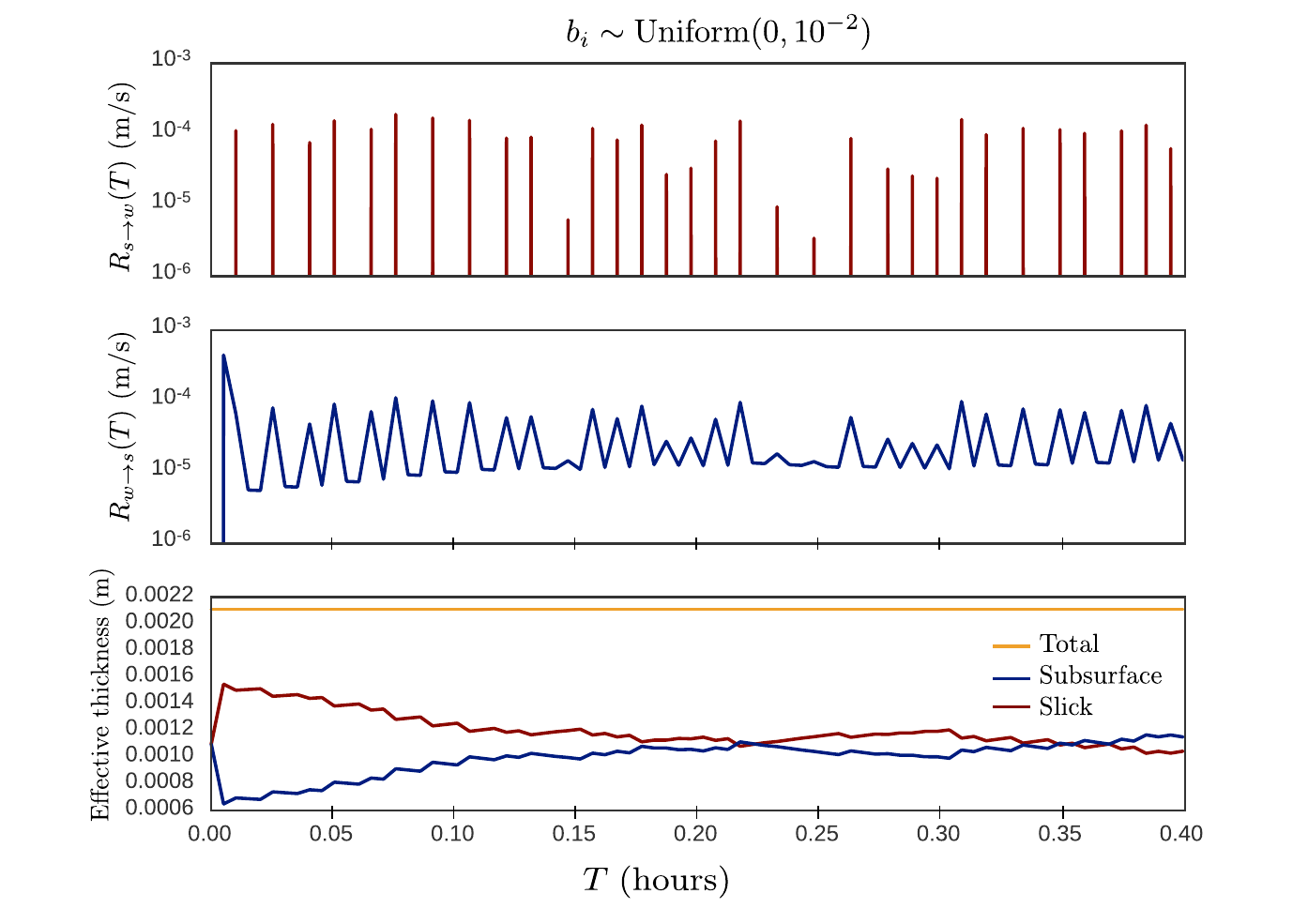}
 \caption{Slick and subsurface oil mass exchanges expressed in terms of effective thickness (volume of oil per unit surface area of ocean) for a wave breaking strength of $b_i \sim \text{Unif}(0,10^{-2})$. See equations (\ref{mastern}) and (\ref{shalfstep}). Top figure: rate of exchange from slick to subsurface $ R_{s\to w}(T) = \sum_{i=1}^{N_b(T)} s(T_i) b_i \delta_{T_i}(T) $. Middle figure: rate of exchange from subsurface to slick $ R_{w\to s}(T) = \sum_{j=1}^{m}  \frac{w_b(r_j)}{H} \bar{n}_j(T)$. Bottom figure: total effective thickness of oil in the slick and subsurface, and their sum (which is shown to be conserved). }
 \label{stack01}
 \end{figure}

\begin{figure}[hbt]
 \centering
\includegraphics[scale=0.6]{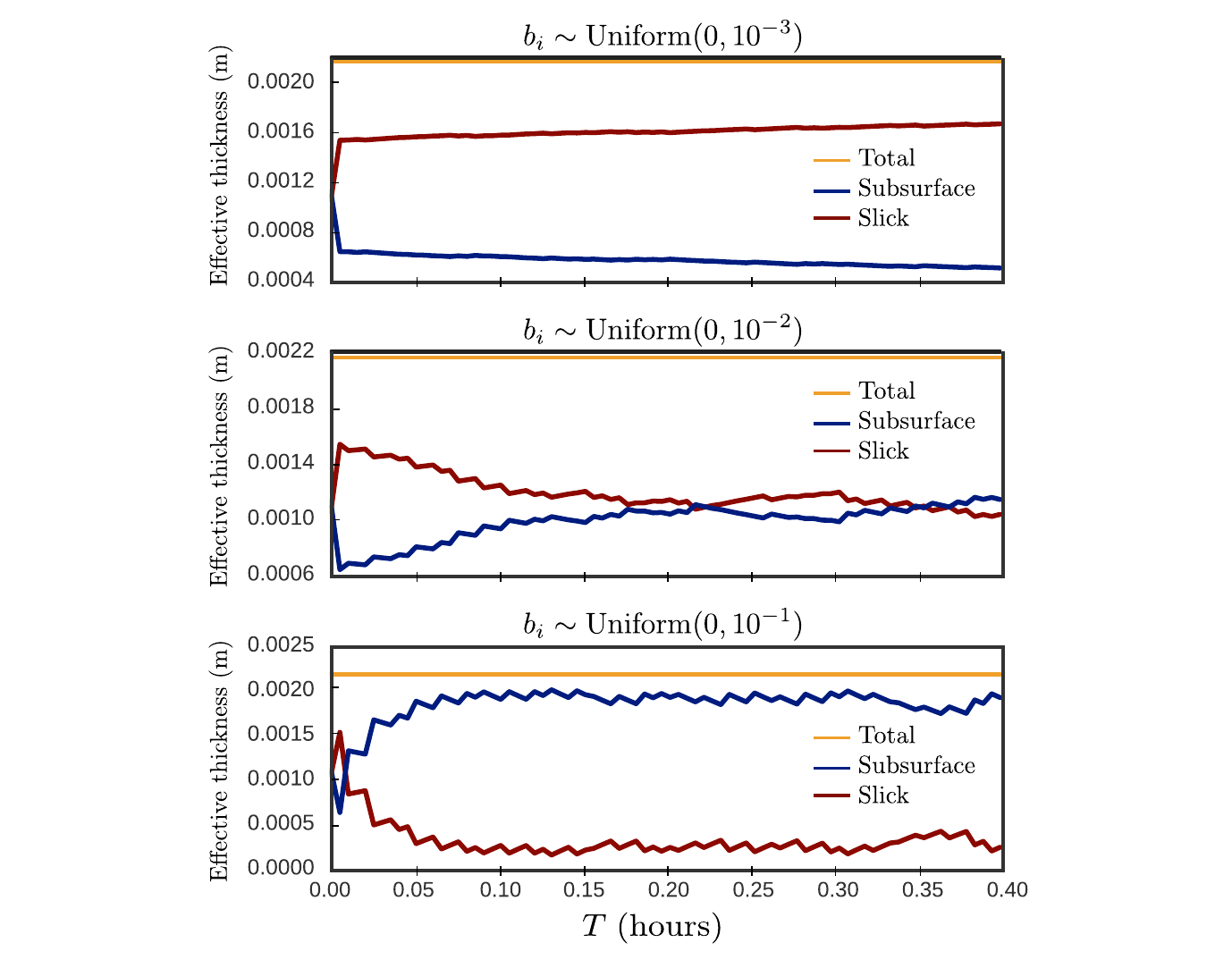}
 \caption{Total effective thickness of oil in the slick and subsurface as a function of time for different
 wave breaking strengths. The dynamics are given by equation (\ref{mastern}) and (\ref{shalfstep}). }
 \label{stackallb}
 \end{figure}

\begin{figure}[hbt]
 \centering
\includegraphics[scale=0.8]{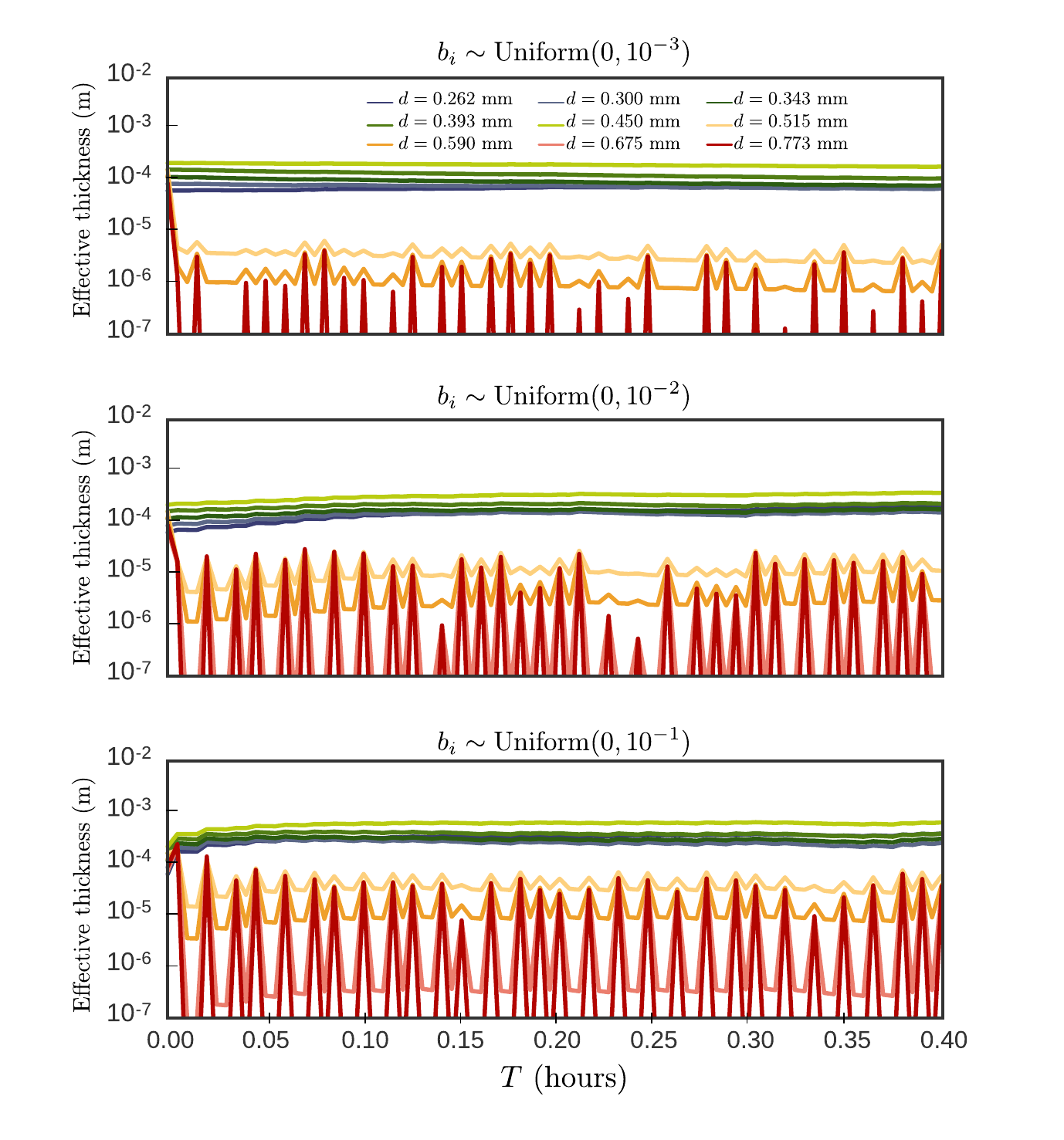}
 \caption{Slick and subsurface oil. Subsurface oil size distributions as a function of time. See Figure \ref{stackallb}.
 For the   wave breaking strengths,  $b_i$= 0.001, 0.01, and 0.1. The  size distribution is $d$.
  Dynamics are given by (\ref{mastern}) and (\ref{shalfstep}). Wave amplitude 
 is 1m, wave period is $T_w= 9.1$s, and the breaking  rate is  $1/6T_w$.  We used 9 size classes. }
 \label{stackallbsize}
 \end{figure}

Figure \ref{stack01} shows the total oil in the slick and the subsurface as a function of time for wave breaking strength  $b_i \sim \text{Unif}(0,10^{-2})$.  The wave breaking strength takes high enough values to force a substantial portion of the oil in the slick to go to the subsurface. The fluxes associated with wave breaking nearly balance, in this case, with the fluxes associated with buoyancy. Larger  $b_i$ lead to wave breaking fluxes dominating.
 The situation changes for a  sufficiently small wave breaking strength. See Figure \ref{stackallb}.
 In this figure we plot the mass balance as a function of time for different wave breaking  strengths $b_i$ of maximum possible value 
 0.001, 0.01, and 0.1. For small $b_i$ the buoyancy fluxes dominate and the mass changes smoothly after a very short transient. As the breaking strength increases, the breaking fluxes dominate, and for very large breaking strengths the mass reaches a quasi-equilibrium configuration.
In Figure \ref{stackallbsize} we plot the equivalent thickness of each size class as it evolves driven by the wave breaking regimes of Figure \ref{stackallb}. The main conclusion drawn from this calculation is that the oil in the large size classes changes the overall balance of oil in the slick and the subsurface, while the small size classes vary in  time more dramatically for any breaking strength.

When transversely coupled stacks are considered, we can introduce advection and dispersion mechanics, which along with transversal heterogeneity of the wave forcing, would alter considerably the dynamics of coupled stacks. We will highlight next a special environmentally-inspired case in which we can explore the stack dynamics coupled with advection and dispersion.

\subsection{Mass Exchanges In The Nearshore Sticky Waters Problem}
\label{sticky}

In RVD we proposed a conceptual model for the slowing down and possible parking of buoyant pollutants, such as oil,  approaching a shore due to ocean circulation. When oil is brought toward the shore by ocean currents, it was found that the advective effects and the dispersive effects can balance each other in this nearshore so as to  slow down or  park a buoyant tracer.  We coined the phrase {\it nearshore sticky waters}  to refer to the phenomenon of slowing or parking of pollutants. The basic mechanics are proposed conceptually in RVD. In RRV the nearshore sticky waters problem  was revisited in order to examine how the shallow-water oil transport model that is being developed here captures the  phenomena. 

Here we incorporate the mass exchange mechanism proposed in earlier sections of this paper into the shallow water oil fate model.   The scenario is presented in Section 6.1 of RRV, and we are only going to point out in what ways the present calculation differs from the ones in RRV. 

\begin{figure}[hbt]
 \centering
(a)\includegraphics[scale=0.25]{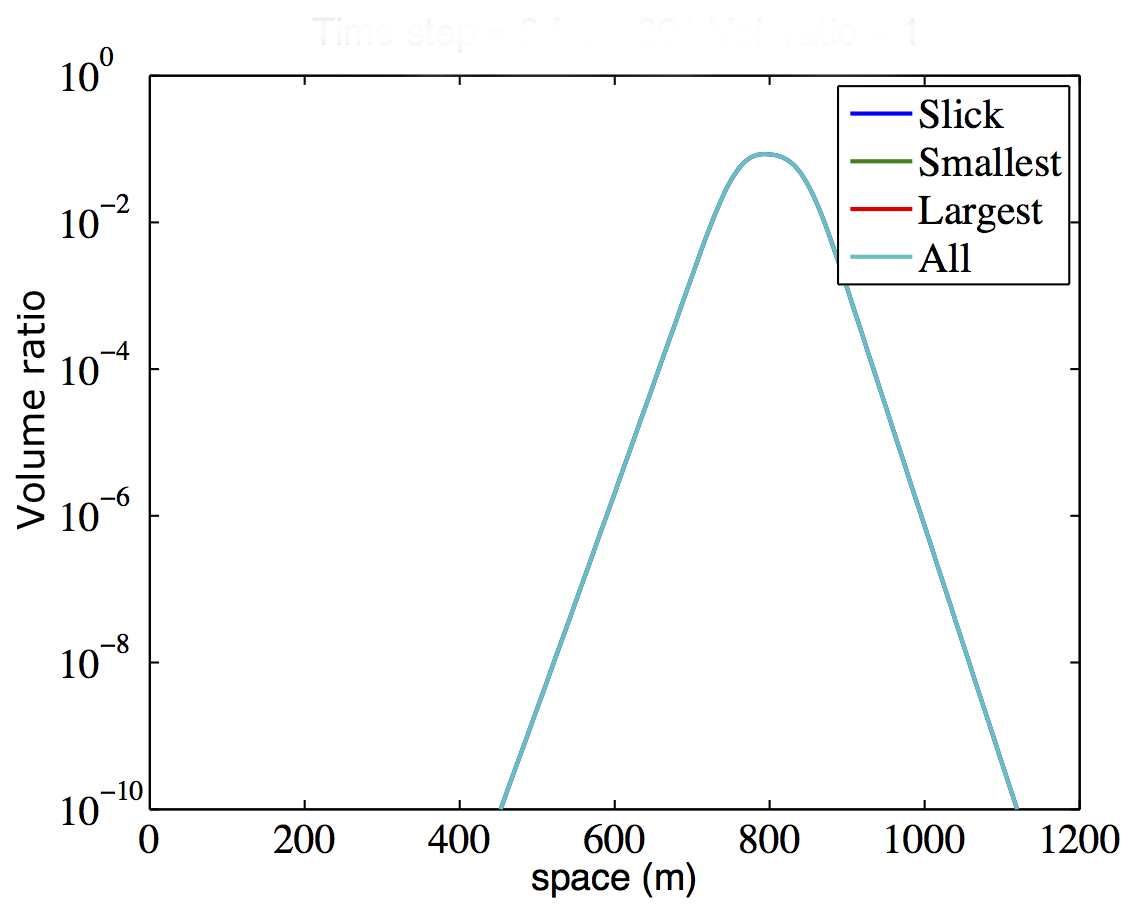}
(b)\includegraphics[scale=0.25]{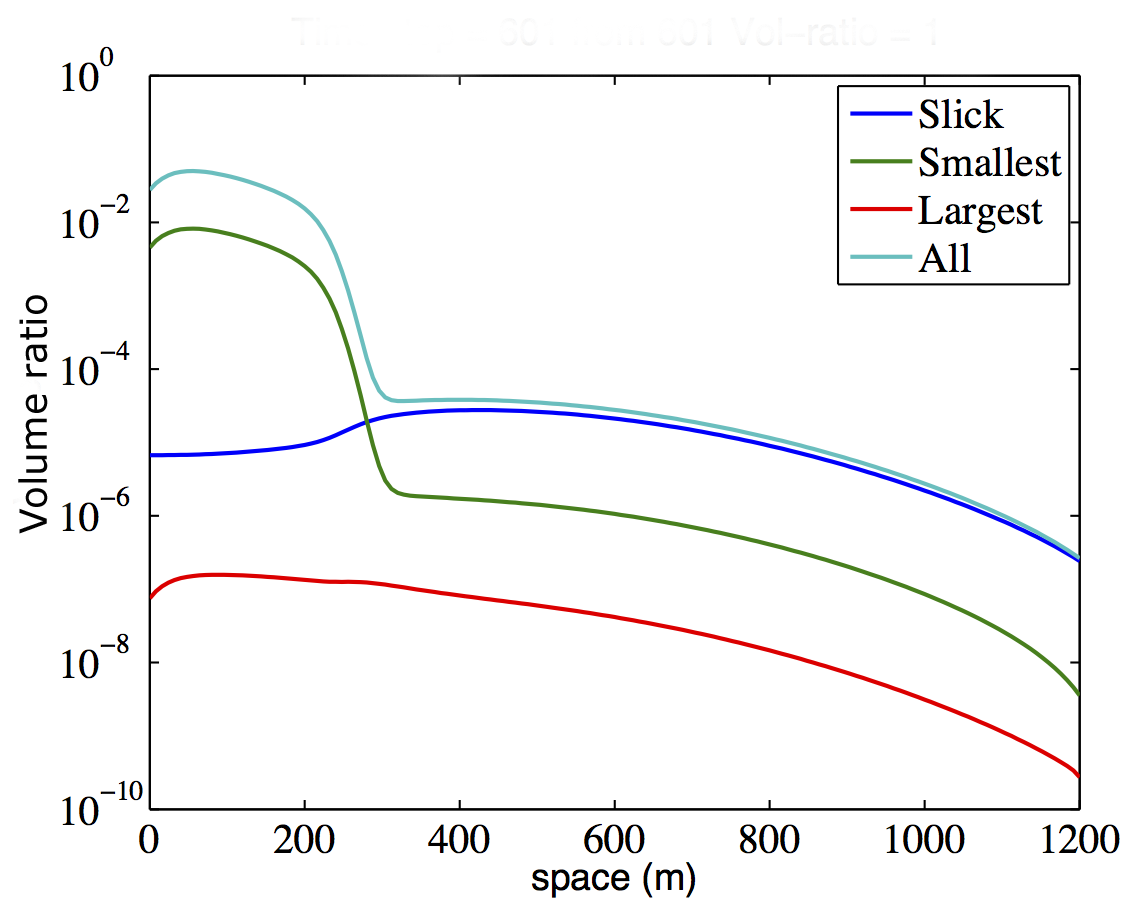}
(c)\includegraphics[scale=0.25]{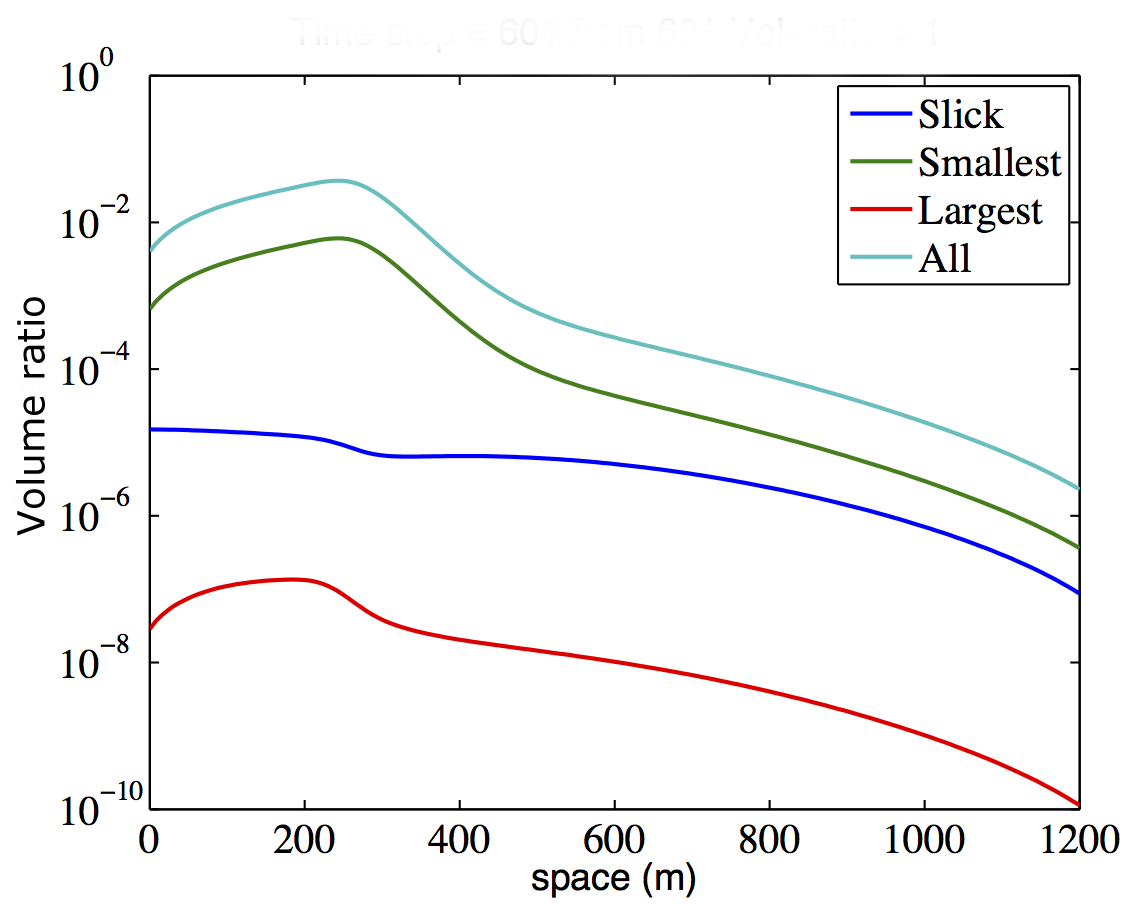}
 \caption{Slick and subsurface oil, initially (a), and after roughly 1000 hours, (b)-(c). The initial oil distribution is entirely contained in the slick. 
The total oil initially and throughout the calculation is $v_{tot}:= 2.334 \times 10 ^{-3}$ m$^3$/m/m;
 (b) corresponds to the oil distribution at the final time, with a mixed layer thickness of $P=1$m, and (c) corresponds to the oil distribution at the final time for  a mixed layer thickness $P=6$m. When the thickness of the mixed layer is large the center of mass of the oil shifts away from the shore. Note that the vertical axis, corresponding to the ratio of oil in the size class or in the slick to $v_{tot}$,  is logarithmic.  }
 \label{stickycase}
 \end{figure}

The changes to the original model, given by Equations (39),  (40), (43) in RRV, and the one captured by  (\ref{seq}), (\ref{ceqfinal}) and (\ref{req}), in the algorithmic process described in Section \ref{themodel}, are as follows:
\begin{itemize}
\item For simplicity, we will use a constant mixed layer depth $P$ in the calculations;
\item the subsurface oil will be captured by 4 size classes, chosen in the range  $r_{1} = 0.0001$m and  $r_{m} = 0.00038$m. The only size class that can buoyantly reach the slick is the oil contained in the largest subsurface size class. In this regard we are following TC, who report that only oil droplets larger than about $0.0005$m are affected significantly by buoyancy forces.
\item  The mass exchange  term in the dynamics of the slick and the subsurface oil, used in the older paper (which reads $\frac{(1-\gamma) s - \gamma P S}{\varsigma(x)}$), are replaced by the mass exchange terms derived in this paper; further,
\item to the mass exchange component of the subsurface (\ref{mastern}) associated with oil entrainment due to wave breaking $\sum_{i=1}^{N_b(T)} N_j s(T_i) b_i \delta_{T_i}(T)$, we add $s(T) \frac{2 \epsilon}{\rho_w g A^2}$. Here, $\epsilon$ is the wave action damping (see Equation (33) of RRV, $\rho_w$ is the ocean density, $g$ is gravity, $A$ is the wave amplitude. In the calculations that follow the wave amplitude in the deeper end of the domain is 1 m, and the wave period is 9.1 s.  The new term accounts for the intense turbulence associated with the shoaling process of the waves. In general it tends to dominate the mass exchanges due to wave breaking in the shoaling region;
\item similarly, to the mass exchange component of the slick (\ref{shalfstep}) associated with oil fluxes to the subsurface,  $ \sum_{i=1}^{N_b(T)} s(T_i) b_i \delta_{T_i}(T)$, we add  $ s(T) \frac{2 \epsilon}{\rho_w g A^2} $ to account for vertical oil fluxes due to wave shoaling;
\end{itemize}
Each computational grid location in the transverse coordinate $x$ has a stack composed of a multi-size subsurface compartment, and an upper slick compartment. Variations along the direction $ y $ of the shore, are ignored. The calculation is performed as follows: the model performs mass exchange calculations using an adiabatic approximation in the transverse distribution of oil among the many coupled stacks.  The mass exchanges within each stack is run for about 1.5 hrs. Then the advection and dispersion across the coupled stacks is performed. This process is repeated. The total time lapsed is in the order of 1000 hrs.
The parameters used in the calculation are the same as those that were used to generate  Figure 7 of RRV.
Figure \ref{stickycase}a shows the initial oil distribution. The total amount of oil is  $v_{tot}$= 2.334 $\times 10 ^{-3}$ m$^3$/m/m, and all of the oil is initially the slick. Figures \ref{stickycase}b-c show the final oil distribution, roughly after 1000 hrs. The oil is reported on a logarithmic scale as the volume ratio with respect to $v_{tot}$ in a size class or in the slick.
 
The cases shown in Figure \ref{stickycase}b-c differ only in the thickness of the mixed layer, 
  with (b) corresponding to $P=1$m, and (c) to $P=6$m. 
 As was shown in RVD and RRV  the $P=1$m case shows little evidence of slowing or parking. In the earlier papers we reported that  
 the oil that initially started in the slick stayed in the slick even very close to the shore. This is no longer the case with the mass exchange
 term proposed in this paper. The slick still contains most of the oil away from the shore, but there is a redistribution of the oil between the slick and the subsurface due to shoaling. The $P=6$m shows that there is a parking of the oil away from the shore: the center of mass is roughly 200m away from the shore. The oil has also redistributed between the slick and the subsurface both close as well as far away from the coast. In both of these cases the bulk of the subsurface oil is found in the smaller droplet size class. 
\section{Summary}
\label{conc}

In this paper and in RRV we propose an Eulerian, mass-conserving, semi-empirical, oil fate model that is specialized to shallow water environments. 
The model, whose development is far from complete at this stage,  is being designed to  capture oil spills at very large spatio-temporal scales required for environmental applications, in the shallow reaches of the ocean. Achieving the large scales is 
 largely due to our ability to parametrize the vertical distribution in terms of depth-averaged quantitites. In RRV we focused
on a proposal for the horizontal transport through advection and the dispersion mechanics. In this paper we focused on the mass exchanges between the slick, which is
formally the oil present on the surface, and the subsurface oil. The key to modeling vertical exchanges between surface and subsurface oil is that
microscale mechanics are not ignored. However, in order to handle large scale dynamics  we cannot resolve micromechanics. In the vertical direction the waves and the turbulence in the oceanic water column, particularly close to the surface, can send oil to the deeper reaches of the water column. At the same time, most oil is buoyant and this buoyancy counteracts the downward flux due to waves and turbulence. Surface tension, viscous effects, and shearing easily breaks up oil into tiny droplets. Stokes balances on droplets tend to keep smaller droplets from resurfacing. Furthermore, oil droplets in high concentrations will merge/coalesce and breakup due to fluid stretching and shearing, affecting the overall distribution of oil droplets. 
This means that  advection, dispersion, droplet dynamics, buoyancy and near surface eddy and shearing forces are critical components of an oceanic oil fate model. In this and in RRV we have proposed models for these phenomena. Other effects, such as reaction, sedimentation, and biodegradation will be added to this model in the future. These bring into the model the inherent challenges associated with handling the dynamics of a buoyant material made up of thousands of simpler chemical components. 
Within the model development program proposed in RRV we will next proceed to incorporate weathering or aging. Oil is a chemically-reacting substance and it is composed of a large number of chemicals.  Clearly, tracking many chemicals increases substantially the computational expense of a general oil transport model. The main thrust of tackling oil aging is thus to derive equations for  a few lumped chemical complexes that mimic the behavior of the thousands of chemicals present in real oil. 

Neither the mass exchange model we propose in this paper and the subsequent weathering model will be specific to our transport model. Hence, in principle, the outcome of these modeling efforts will have a bearing and perhaps find application in other 
oil-fate models.

The mass exchange model proposed in this study upscales the microscale mechanics and captures merging, splitting, vertical transport due to buoyancy and turbulence. The exchanges preserve mass conservation, a feature we consider  essential to an  oil-fate model. The upscaling is achieved by  exploiting the disparate time scales of (slower) advection/dispersion and (faster) microscale mechanics. This assumption has obvious computational implications: the faster microscale mechanics can be upscaled to the slower scales of transverse oil motion. Properly exploited, this leads to a model that can handle large transverse spatio-temporal scales without resorting to small scale stationarity that could allow for the use of homogeneization of  the small scale dynamics.  

The merging and coalescence mechanics in the model were benchmarked against data and shown to have reasonable consistency. When compared with our intuition, the vertical fluxes due to wave breaking and buoyancy were shown to be reasonably well parametrized with minimal free parameters and with enough phenomenology to capture general oceanic situations. Specialized experiments and field campaigns would be necessary to test and refine the model further.

In this study we also applied the mass exchange model to a conceptual model that explains how oil and other buoyant pollutants behave dynamically in the near shore zone. Nearshore sticky waters still persist, as proposed in RVD: the added micromechanics and mass exchanges do not change the basic premise behind the suggested explanation for the slowing down or parking of oil and other buoyant pollutants that can reach the shores via oceanic circulation.
  The mass exchange model suggests that the mixed layer depth has a significant effect on the overall distribution of oil: when the mixed layer is thin the distribution remains fairly constant in time, far from the shore. However, this is not the case when the mixed layer is large.  Our model suggests that oil droplet distributions play a critical role in the dynamics of nearshore oil dynamics, and a more careful and extensive analysis of the model and of field data merits attention.

\section*{Acknowledgements}

This work was supported by GoMRI and by NSF DMS  grants 1524241, 1434198 and 1109856. JMR also acknowledges
 the hospitality of  Stockholm University, where some of this work was done, and the support provided by a \
 Stockholm University Rossby Fellowship.

\bibliography{swwe,oil}

\end{document}